# Multi-dimensional sparse time series: feature extraction

Marco Franciosi, Giulia Menconi
Dipartimento di Matematica Applicata, Università di Pisa

*Abstract*— We show an analysis of multi-dimensional time series via entropy and statistical linguistic techniques. We define three markers encoding the behavior of the series, after it has been translated into a multi-dimensional symbolic sequence. The *leading component* and the *trend* of the series with respect to a mobile window analysis result from the entropy analysis and label the dynamical evolution of the series. The *diversification* formalizes the differentiation in the use of recurrent patterns, from a Zipf law point of view. These markers are the starting point of further analysis such as classification or clustering of large database of multi-dimensional time series, prediction of future behavior and attribution of new data. We also present an application to economic data. We deal with measurements of money investments of some business companies in advertising market for different media sources.

## I. Introduction

In the last decades of twentieth century, several methods from nonlinear dynamics have been proposed to analyze the structure of symbolic sequences. Different statistical methods have been introduced to characterize the distribution of words, or combinations of symbols, within the sequences, and many applications (e.g. to DNA analysis) has been found.

One of the most significant is based on an asymptotic measure of the density of the information content. In an experimental setting, information content may be approximated by means of compression algorithms (see for instance [1]). The notion of information content of a finite string can be used also to face the problem of giving a notion of randomness. Namely, this leads to the notion of *entropy* $h(\sigma)$ of a finite string $\sigma$, which is a number that yields a measurement of the complexity of $\sigma$ (see Section II for details). Intuitively, the greater the entropy of a string, the higher its randomness in the sense that it is is poorly compressible.

Another useful tool is given by statistical linguistic techniques such as the Zipf scaling law, which offers a nice methodology that can be applied in order to characterize specific aggregation patterns or to identify different "languages". The so-called Zipf analysis [10] is useful to understand how variable the distribution of patterns is within a symbol sequence. The basic idea is that the more variable the observed sequences are, the more variable the measurements and the more complex the obtained language.

These techniques can be applied to some time series

$$X = (x_1 x_2 \ldots x_t)$$

by considering a translation into a finite symbol sequence, usually given by means of a uniform partition of its range (see Section II for details). In this way one can use statistical linguistic techniques in the analysis of series $X$ trying to find some *aggregation patterns* or *global scores* allowing feature and marker extraction, useful to label the series in view of classification, clustering or attribution.

The purpose of this paper is showing an application of this approach in order to analyze multi-dimensional time series.

By multi-dimensional time series we mean a finite set

$$\mathbb{X} = \left( \begin{array}{c} X_1 \\ \vdots \\ X_N \end{array} \right)$$

of data, where each data $X_j$ is a finite array of real numbers coming from subsequent discrete measures of some empirical phenomena (e.g., weekly data): $X_j = (x_{j,1} \ldots x_{j,t})$.

Multi-dimensional time series appear if one deals with multiple measurements on some objects/phenomena, each one focused on some *a priori* structure of the process under study. Examples of such multimedia mining are given when considering different measurements (such as temperature, pressure, velocity) of the same physical phenomenon or taking different clinical data (such as pulse-rate, blood pressure, oxygen saturation, etc...) of one single patient (see Ref. [7]). Other examples appear analyzing financial markets, where it is worth to look at different behaviors of one company, e.g., in order to define some good strategy. From this point of view it is particularly interesting the *Advertising Market*, where it is natural to consider money investments of some business companies for different media sources.

Frequently, experimental (one-dimensional) time series are short and no longer prolongable. Moreover, they may be *sparse* in the sense that the measurements they come from are not homogeneous in time and many values are null due to a failure in data acquisition or (e.g. when some investments are recorded) at some time step there is nothing to be measured; that is, they come from sparsely sampled, incomplete or noisy data. More formally, a time series is sparse when, in the range of $\tau$ time steps, the null measurements are at least $\tau\delta$. That is, in the series the density of null values $N(X_j) \doteq \#\{x_{i,j} = 0 \, : \, i=1,\ldots,t\} \geqslant t\delta$ where for instance $\delta \sim \frac{1}{4}$. Actually, first one has to discriminate whether some event occurrs or not, then its extent. Not rarely, statistical methods to analyze time series only take into consideration the magnitude of realization of some event, neglecting the case when there are no events (e.g. cumulative random walks), while the data time aggregation may be a discriminant feature by itself.

The advantage of dealing with a multi-dimensional time series is that, on the one hand, it offers a global point of view and shows some critical pathologies arising from evident discrepancies, whereas, on the other hand, it permits to integrate the information contained in each one-dimensional time series of $\mathbb{X}$ and therefore

Dipartimento di Matematica Applicata, Università di Pisa, Via Buonarroti 1C, I-56127 Pisa, Italy. Correspoding e-mail: menconi@mail.dm.unipi.it

it is useful when each array is sparse and short.

Following this line, entropy and Zipf analysis can be applied to each array of a given multi-dimensional time series $\mathbb{X}$, allowing a global perspective of $\mathbb{X}$ to be achieved. This analysis is particularly useful when arrays $X_j$ are pairwise incomparable (e.g., they represent different physical measures of some phenomenon) or if the the values acquired in different series are different in magnitude (i.e. given $X_j$ and $X_h$, it is $x_{j,i} \gg x_{h,i}$ for every $i$). In this paper we show how to label multi-dimensional time series by means of a few markers resulting from entropy analysis and Zipf linguistic statistics. Such markers by themselves are a simple way to characterize the dynamical structure of the phenomenon under analysis. Furthermore, they may be used to create new customized methods of clustering and feature attribution.

To illustrate our method we present here an application of these techniques to economic data coming from advertising market. In our example we shall deal with measurements of money investments of some business companies in advertising market for different media sources (TV, radio, newspapers, etc). Nevertheless, the features describing some company and expressing its typical traits of investment policies may come from other measures, derived from the integration of the results of entropy analysis for each component. Such features may be used to understand the behavior of each company with respect to the different media.

*A. Notations*

Throughout this paper, we shall use the following notations for time series:
- $X$: one-dimensional time series
- $\mathbb{X}$: multi-dimensional time series
- $\sigma, S$: finite symbolic sequence
- $\mathbb{S}$: multi-dimensional symbolic sequence

## II. SYMBOLIC ONE-DIMENSIONAL TIME SERIES

Consider a one-dimensional time series $X = (x_1 x_2 \ldots x_t)$ of length $t$. In a standard way, we translate $X$ into a finite symbol sequence $S$ by means of a uniform partition of its range, as follows.

Fix a positive integer $L$ to be the size of some alphabet $\mathcal{A} = \{1, 2, \ldots, L\}$ and let $I_1 \doteq \min\{x_1, \ldots, x_t\}$ and $I_{L+1} \doteq \max\{x_1, \ldots, x_t\}$. Then divide the interval $[I_1, I_{L+1}]$ into $L$ uniformly distributed subintervals.

To each value $x_i$ in $X$ we associate symbol $\ell \in \mathcal{A}$ iff $x_i \in I_\ell$. We obtain a sequence $S$ which is the symbolic translation of series $X$.

Symbolic sequence $S$ may be considered as a phrase written in some language. The more variable the observed data, the more complex the obtained language. Entropy is a way to characterize the way the phrase $S$ is built, while Zipf analysis refers to the typical recurrent words.

*A. Entropy*

One of the most significant tools from the modern theory of nonlinear dynamics used to analyze time series of biological origin is related to the notion of *information content* of a finite sequences as introduced by Shannon in [9]. The intuitive notion of information content of a finite word can be stated as "the length of the shortest message from which it is possible to reconstruct the original word" and a formal mathematical definition of this notion has been introduced by Kolmogorov using the notion of universal Turing machine (see [5]). We will not enter into the details of the mathematical definition, but simply use the intuitive notion of information content we stated.

The method we use to study the information content of a finite sequence is related to *compression algorithms*. The compression of a finite sequence reflects the intuitive meaning of its information content.

Let $\sigma = (s_1 s_2 \ldots s_t)$ be a $t$-long sequence written in the finite alphabet $\mathcal{A}$. Let $\mathcal{A}^t$ be the set of $t$-long sequences written using $\mathcal{A}$ and let the space of finite sequences be denoted by $\mathcal{A}^* := \cup_t \mathcal{A}^t$.

A *compression algorithm* on a sequence space is any injective function $Z : \mathcal{A}^* \to \{0, 1\}^*$, that is a binary coding of the finite sequences written on $\mathcal{A}$.

The information content of a word $\sigma$ w.r.t. $Z$ is the binary length of $Z(\sigma)$, the compressed version of $\sigma$. Hence

$$I(\sigma) \doteq \text{Information Content of } \sigma = |Z(\sigma)|$$

The notion of information content of a finite string can be used also to face the problem of giving a notion of randomness. Namely, we can think a string to be more random as less efficient is the compression achieved by a compression algorithm. This leads to the notion of *entropy $h(\sigma)$ of a finite string*, defined as the compression ratio (i.e. the information content per unit length):

$$h(\sigma) \doteq \text{Entropy of } \sigma = \frac{I(\sigma)}{|\sigma|} = \frac{|Z(\sigma)|}{t}$$

It holds that $0 < h(\sigma) \leqslant 1$ and moreover the greater the entropy of a string, the higher its randomness in the sense that it is is poorly compressible.

*Remark 1:* When analysing a symbolic string with entropy tools, it is convenient to consider asymptotic properties, hence assuming to have an infinite stationary[1] sequence. We can make this assumption to obtain some mathematical results on the complexity of a string. For an infinite sequence $\tilde{\sigma} = (s_i)_{i \geqslant 1}$ written on an alphabet of size $L$, we can define the asymptotic compression ratio $K(\tilde{\sigma}; L) \doteq \lim_{n \to \infty} h((s_1, \ldots, s_n))$. If we are dealing with symbolic translations of some time series $Y$ being the (infinite) orbit of a dynamical system then we may consider partitions of increasing length $L$, such obtaining an infinite set of symbolic translations of $Y$. For each size $L$, we have $\tilde{\sigma}(L)$ and obtain $K(Y; L) \doteq K(\tilde{\sigma}(L); L)$. In this setting, the limit $\lim_{L \to \infty} K(Y; L)$ is the metric entropy of the dynamical system (see Ref.[2]).

*Remark 2:* Even in the case of finite sequence $\sigma$, the property of being stationary allows a proper connection of the entropy of $\sigma$ to the above mentioned theory to be established. Since an experimental time series $Y = (y_1, \ldots, y_{t+1})$ is hardly stationary, a way to make it close to be stationary is to consider the difference series $D = (d_1, \ldots, d_t)$ where $d_j \doteq y_{j+1} - y_j$ and to apply the symbolic analysis to that $D$. Again, in the infinite case, the entropy of $Y$ and $D$ coincide and this motivates the use of $D$ also in the finite case.

---

[1]An infinite sequence $Y = (y_i)_{i \geqslant 1}$ is stationary if for each $k \geqslant 1$ and for each $k$-long finite sequence $\alpha = a_1 \cdots a_k$ the $Prob\{(y_i \cdots y_{i+k-1}) = \alpha\}$ is independent of $i$.

## B. Linguistic analysis

Some time series $X$ may be read as a sequence of measurements governed by some dynamic rules driving the time change in the measured values. Notwithstanding the entropy measures the rate of variability in the series, other crucial hints about the series may come from statistical analysis of the patterns described by the series, as words in a language generating the symbolic string associated to the series. Thus, we performed the so-called Zipf analysis [10], useful to understand how variable is the distribution of patterns within a symbol sequence.

Given a finite symbol sequence $\sigma$ of length $t$, let us fix a word size $p < t$ and let us consider the frequency of all words of length $p$ within $\sigma$. Let us order such words according to their decreasing frequency. This way, each word has a rank $r \geqslant 1$. The Zipf scaling principle asserts that in natural languages the frequency of occurrence $f(r)$ of word of rank $r$ is s.t. $f(r) \sim (1/r)^\rho$ where $\rho \leqslant 0$. In an experimental setting, the value of Zipf coefficient $\rho$ may be calculated via linear regression on the frequency/rank values in bilogarithmic scale. A low scaling coefficient is connected to high variability of words: were the words uniformly distributed, the scaling coefficient would be zero. Thus, the more variable the observed sequences are, the more complex the obtained language is and the more variable the measurements are. The most famous example of Zipf's law is the frequency of English words. Anyway this kind of rank-ordering statistics of extreme events, originally created to study natural and artificial languages, had interesting applications in a great variety of domains, from biology [6], to computer science [3], to signal processing [4] and to meteorology [8] (this list may not be exhaustive).

## III. MULTI-DIMENSIONAL TIME SERIES

In this section we show how to extend the above mentioned tools to multi-dimensional time series. We may assume that the one-dimensional series have comparable length. We do not require them to have the same length $t$, but we require that each length $t_1, \ldots, t_N$ are of the same order $t$ and all the measurements refer to the same time lag of observation of the phenomenon. This discrepancy may be overcome by adding null values when there is lack of measurements, when this does not affect the sense of the analysis.

Given an alphabet size $L$, we associate to each multi-dimensional time series $\mathbb{X} = (X_1, \ldots X_N)^T$ a multi-dimensional symbolic sequence $\mathbb{S} \doteq (S_1, \ldots S_N)^T$ where $S_j$ is the symbolic sequence associated to the one-dimensional sequence $X_j$.

## A. Global Entropy

Given $\mathbb{X}$ and its symbolic translation $\mathbb{S} = (S_1, \ldots S_N)^T$, we can compute the entropy of each component and obtain the entropy vector:

$$H(\mathbb{X}) = \begin{pmatrix} h(S_1) \\ \vdots \\ h(S_N) \end{pmatrix} \quad (1)$$

Natural measures that may be taken under consideration are the Euclidean norm and the $\ell_1$ norm of $H(\mathbb{X})$:

$$||H(\mathbb{X})|| = \sqrt{\sum_{i=1}^{N}[h(S_i)]^2} \quad (2)$$

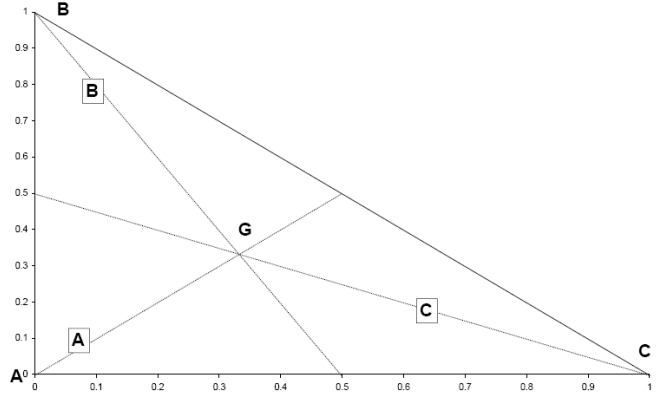

Fig. 1

*Symplex $\Delta^3$ in $\mathbb{R}^2$ and influence areas of vertices A,B,C w.r.t. centroid G.*

$$||H(\mathbb{X})||_1 = \sum_{i=1}^{N} h(S_i) \quad (3)$$

They quantify the extent of global entropy over all the components describing the process $\mathbb{X}$.

Notice that the vector $H(\mathbb{X})$ yields a simple way to characterize the behavior of the series $\mathbb{X}$ and it is not uncommon to see that symbolic series associated to experimental measurements with different magnitude may have almost the same entropy (for instance, take a series and create a new one just doubling the values of the first one, then in the symbolic model they are the same sequence).

Assume that the entropy vector is not null. For what concerns the role of single components, we may investigate what their relative influence is by means of the following symplex analysis. Choose some component, say the $N$-th.

We consider the $(N-1)$-dimensional symplex

$$\Delta^N = \left\{ \begin{pmatrix} y_1 \\ \vdots \\ y_{N-1} \end{pmatrix} : \sum_{i=1}^{N-1} y_i \leqslant 1 \ and \ y_i \geqslant 0 \ \forall i \right\}$$

and the natural projection $P$ of the vector $H(\mathbb{X})$ onto $\Delta^N$, i.e.

$$P = P(\mathbb{X}) = \begin{pmatrix} \frac{h(S_1)}{||H(\mathbb{X})||_1} \\ \vdots \\ \frac{h(S_{N-1})}{||H(\mathbb{X})||_1} \end{pmatrix}$$

The position of the point $P$ w.r.t. the vertices and the centroid $G$ of $\Delta^N$ is a static feature of the process represented by $\mathbb{X}$, showing which one of the $N$ components is leading the dynamics. Indeed, the vertex $V_N = (0, \ldots, 0)^T$ is associated to the $N$-th component, whereas the vertices $V_1 = (1, 0, \ldots, 0)^T$, $V_2(0, 1, 0, \ldots, 0)^T, \ldots V_{N-1} = (0, 0, \ldots, 1)^T$ correspond to the components labeled by $1, 2, \ldots, N-1$.

For each vertex $V_j$, consider the hyperplanes connecting $N-2$ other vertices to the centroid and not containing $V_j$. They partition the symplex $\Delta^N$ into $N$ regions, representing the *influence areas* of each vertex (see an example for $\Delta^3$ in Fig.1). Therefore, if the influence area relative to point $P$ is that of vertex $V_d$, then the dynamics of $\mathbb{X}$ is driven by the *d-th* component (called *leading* component), in the sense that the *d-th* entropy coefficient is prevailing on the others and the dynamic of that component is to be taken under observation more than the others'. We denote by $\mathcal{L}(\mathbb{X})$ the leading component corresponding to the influence area of multi-dimensional time series $\mathbb{X}$.

If we have a collection $\mathcal{B} = \{\mathbb{X}^1, \ldots, \mathbb{X}^b\}$ of $N$-dimensional time series, we can apply the above procedure for every $\mathbb{X}^j$ obtaining a collection of $b$ points $\{P^1, \ldots, P^b\} \subset \Delta^N$. This way, one can see first the position w.r.t. the centroid $G$, second the neighborhood relations, showing influence areas and common behaviors. For an explicit example we refer to Section IV.

*B. Entropy walk*

A dynamic feature showing the trend of entropy production of series $\mathbb{X}$ may be extracted by a mobile window entropy analysis, as follows.

Consider some multi-dimensional series $\mathbb{X} = (X_1, \ldots, X_N)^T$. Let $t$ be the length of each component time series. Fix $k$ be some positive integer. From each series $S_j$ ($j = 1, \ldots, N$) within $\mathbb{S}$, the symbolic model of $\mathbb{X}$, we may extract $k$ subseries $W_1, \ldots, W_k$ in many ways: for instance, overlapping windows, non-overlapping windows, random starting points (fixed once for each collection $\mathcal{B}$ of multi-dimensional time series), etc. We only require all the $k$ subseries have the same length; this implies that the choice of $k$ should keep the length of the subseries sufficiently long for the entropy analysis to be meaningful. For each window we calculate the entropy. We repeat the same for every series in $\mathbb{X}$. We obtain a matrix of entropy vectors in $[0,1]^{N \times k}$, the moving vector of $\mathbb{X}$ denoted by $\mathcal{M}(\mathbb{X})$ whose rows are:

$$\mathcal{M}_1 \doteq \Big(h(W_{1,1}), \ldots, h(W_{1,k})\Big)$$
$$\ldots\ldots\ldots$$
$$\mathcal{M}_N \doteq \Big(h(W_{N,1}), \ldots, h(W_{N,k})\Big)$$

If we are considering a collection $\mathcal{B}$ of multi-dimensional time series, we shall deal with a collection of moving entropy vectors:

$$\mathbb{M}(\mathcal{B}) = \Big(\mathcal{M}(\mathbb{X}^1), \ldots, \mathcal{M}(\mathbb{X}^b)\Big)$$

Again for each index $j = 1, \ldots, b$, we may associate to series $\mathbb{X}^j$ a *sequence* of points in the symplex $\Delta^N$:

$$\mathcal{W}^j = (P_1^j, \ldots, P_k^j) \tag{4}$$

where $P_1^j$ is the point in the symplex corresponding to the entropy vector $(h(W_{1,1}^j), \ldots, h(W_{1,N}^j))^T$, the one relative to the first window, etc.

Please notice that if — in the static context— to each multi-dimensional series in the collection $\mathcal{B}$ just *one* point is associated, in this dynamic context we define an entropy $walk$ relative to each multi-dimensional time series. We study each walk in $\{\mathcal{W}^1, \ldots, \mathcal{W}^b\}$ to characterize the trend of original series in collection $\mathcal{B}$ and to show how to use it to predict future steps of the series as well as to decide whether some new subseries is in accordance with past ones.

The entropy value is a marker of the dynamic change in the time series. The higher is the entropy, the higher is the variability of the series, therefore the more "impredictable" is the future of the series. The entropy walk is a way to look how the entropy changes with time within the sequence. Were the points colinear, the entropy change is balanced and the dynamic change is homogeneous; were the points more scattered, the dynamic rules changed and the process may need a finer observation.

We shall define a trend from which many predictive techniques on the dynamic change of the multi-dimensional time series may be derived.

We calculate $\mathcal{R}$, the linear regression of the points defining the entropy walk $\mathcal{W}$ in $\Delta^N$. The trend of the walk is the pair

$$\mathcal{T}(\mathcal{W}) = (\mathcal{A}, \alpha) \tag{5}$$

where $\mathcal{A}$ is the leading component of the last window and $\alpha$ is the direction of line $\mathcal{R}$ when oriented following the chronological order of the points. The trend itself provides a predictive scenery for the dynamic change in the series.

As a second step, the trend is useful to say whether some new point is in accordance to the past ones. Assume we have a point $Q \in \Delta^N$, say the point associated to some $(k+1)$-th window. We aim at understanding whether it comes from a dynamics in common with the one driving the past walk, that is we aim at verifying how much dynamics the $(k+1)$-th window shares with previous $k$ windows.

There are many different ways to do it; we decided to apply the following criterium:

*If the distance of $Q$ from the linear regression is not greater than the mean distance of points within the entropy walk, then we say that the point $Q$ is within the walk. Otherwise, it is outside the walk.*

Let $(\mathcal{A}, \alpha)$ be the trend of the entropy walk and assume $Q$ is outside the walk. We may apply a second order analysis and use the influence area of the new point as lighter marker of dynamic change: were it different from $\mathcal{A}$, then the process under examination is undergoing an abrupt change. In the case the influence area of $Q$ coincide with the past one, then we may say that the change is still slightly acceptable.

*C. Global Linguistic analysis*

For what concerns multi-dimensional time series, we recall that they are assumed to be short, therefore the statistics is quite poor. Nevertheless, what may be distinctive is the use they do of the distinct words. Moreover, we define a marker of pattern differentiation as follows. Fix once and for all a pattern size $p$ which is sufficiently long w.r.t. the order of the series length $t$. Given a multi-dimensional series $\mathbb{X} = (X_1, \ldots, X_N)$, we calculate the Zipf coefficient for each component $(\rho_1, \ldots, \rho_N)$ and denote by $\mathcal{D}$ the diversification:

$$\mathcal{D}(\mathbb{X}) = 1 + \frac{1}{N} \sum_{j=1}^{N} \rho_j \tag{6}$$

This way, the mean Zipf coefficient gives an estimate of the degree of differentiation in the use of most frequent patterns of length $p$ within series in $\mathbb{X}$. For values of $\mathcal{D}$ close to 1, there is a high diversification of patterns that tend to be used indifferently since their distribution is almost uniform. For values of $\mathcal{D}$ close to 0, the language of the $p$-patterns is rich and there exist some rules giving more importance to some patterns despite others, therefore the distribution of words is no longer balanced. If $\mathcal{D} < 0$ then the words are extremely unbalanced and typically there are a few words used recurring very frequently while most of the words are rarely used.

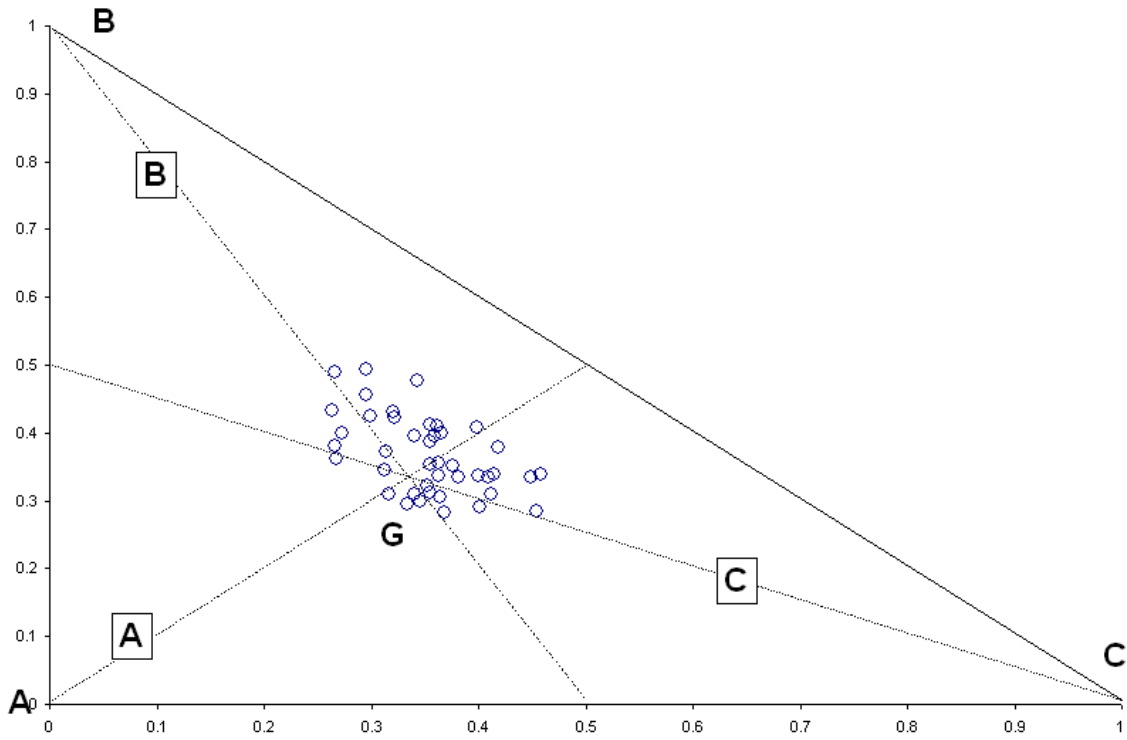

Fig. 2

*Influence areas for the complete multi-dimensional series of 42 brands on the symplex $\Delta^3$ in $\mathbb{R}^2$. Vertex A is relative to radio component, B is relative to magazine component and C is relative to newspaper component.*

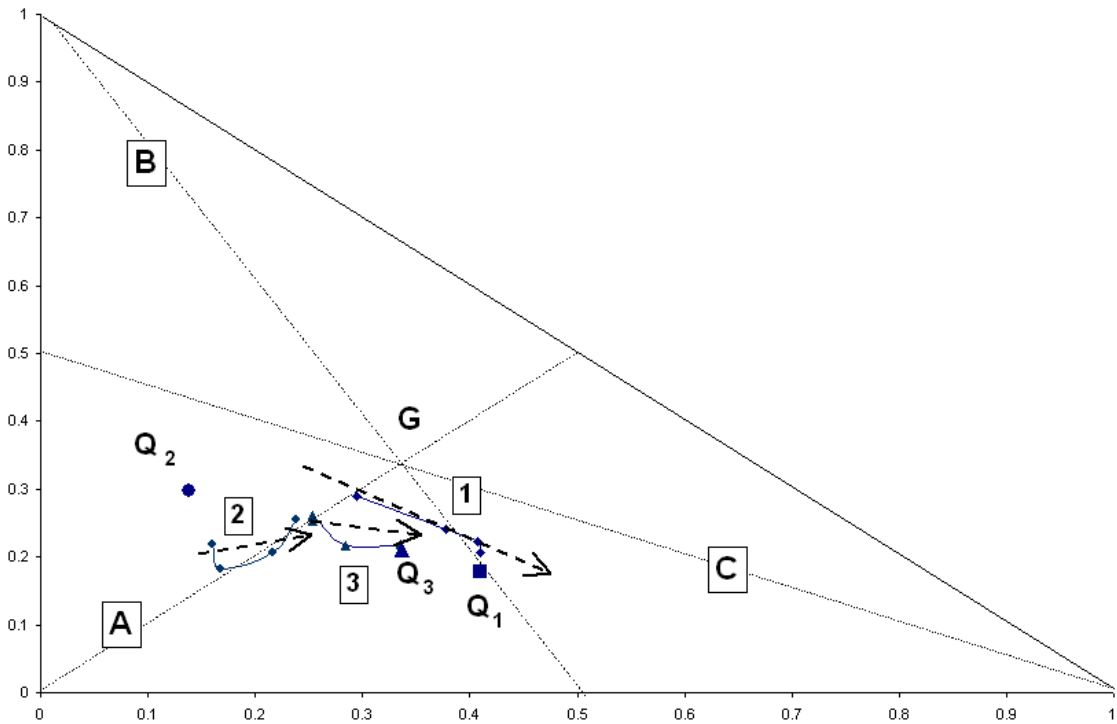

Fig. 3

*Symplex $\Delta^3$ in $\mathbb{R}^2$: entropy walk and trend. Example for three brands $b_1$ (plotted with $\square$), $b_2$ (plotted with $\bigcirc$) and $b_3$ (plotted with $\triangle$) (see text).*

## D. Markers

On conclusion, to each multi-dimensional time series $\mathbb{X}$, we may associate the following markers:

- leading component of the complete series $\mathcal{L}(\mathbb{X})$ as introuced in section III-A
- trend $\mathcal{T}(\mathcal{W})$ w.r.t. $k-$window analysis, following (5)
- diversification $\mathcal{D}(\mathbb{X})$, as defined in (6)

As already discussed, these markers should be the starting point of further analysis such as classification or clustering of large database of multi-dimensional time series, prediction of future behavior and attribution of new data. Finally, let us remark that to the above markers other direct measures may be added, depending on what process we are dealing with. An example is given in the following application section.

## IV. EXPERIMENTAL APPLICATION

We applied our method to 3-dimensional time series related to 42 objects. The data come from Nielsen Media Research data base of weekly investments in advertisement on three Italian media from 1996 to 2006, therefore each object is a brand in the market and each series has 585 non-negative data. The components are the money spent on radio, on magazines and on newspapers, respectively.

The original series were pre-processed in order to make them more stationary; consequently we worked on the difference series, as explained in Remark 2. We applied a symbolic filter with alphabet size $L = 4$ (from abrupt decrement to abrupt increment of investments).

The entropy was calculated using the Lempel-Ziv based algorithm CAST0Re [1]. We recall that any optimal compression algorithm (i.e. on almost every infinite sequences the entropy of the source is reached) may be used.

The series are 3-dimensional, therefore the symplex we use is $\Delta^3 \subset \mathbb{R}^2$ where the vertices are A (relative to radio component), B (relative to magazine component) and C (relative to newspaper component).

The window analysis was exploited over the period 1996-2005 using 4 windows approximately 7-years long (350 measurements) and overlapping for 6 years (first year out, new year in). The measurements concerning year 2006 were used to build another window $Q$ on which the trend was tested as predictive measure (see subsection III).

On Fig. 2, the influence areas of 42 brands are shown. They are almost all close to the barycentre $G$. Nevertheless, their global positioning still suggests that some of them tend to be driven by one specific component.

Three brands $b_1$, $b_2$ and $b_3$ have been considered to exemplify the trend analysis. Fig. 3 shows the entropy walks (solid lines) and the trends (arrows) for brands $b_1$ (plotted with $\square$), $b_2$ (plotted with $\bigcirc$) and $b_3$ (plotted with $\triangle$). Three new points $Q_1 = \square$, $Q_2 = \bigcirc$ and $Q_3 = \triangle$ represent the position in $\Delta^3$ of the subseries that have been tested on whether they are within or outside the respective walk. We deduce that $Q_1$ is outside of brand $b_1$ entropy walk and the leading component also changed, while $Q_2$ is again outside $b_2$ walk, but the leading component remains the same. Finally, $Q_3$ is within the walk.

As a result on the global collection of 42 brands, we obtained 62% within-walk predictions (26 brands), while of the remaining 16 outside-walk brands, only 2 changed leading component.

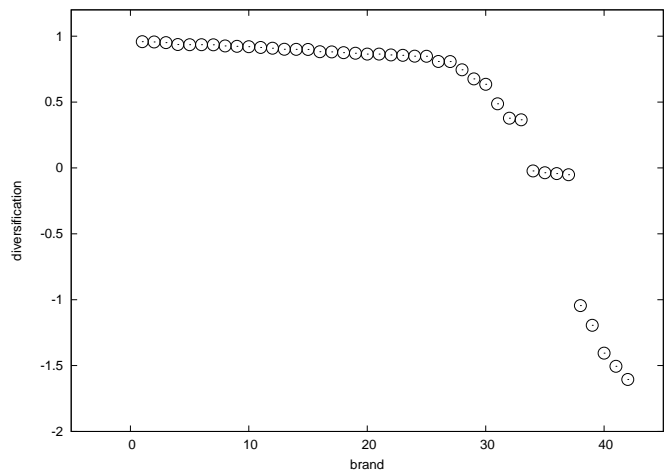

Fig. 4

*Diversification coefficient for 42 brands.*

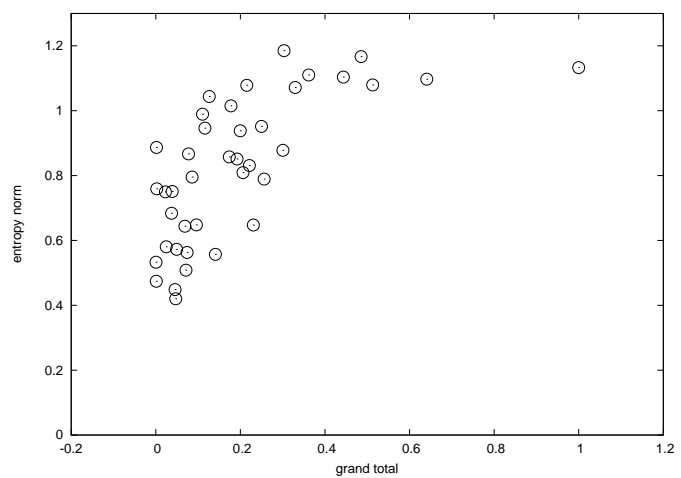

Fig. 5

*X axis: increasing grand total investments (normalized to $[0, 1]$) for the collection of 42 brands. Y Axis: the Euclidean norm of their entropy vectors.*

Other interesting properties of the multi-dimensional series come from the linguistic analysis.

First, some technical details. We exploited Zipf analyis on the symbolic series built on an alphabet with 4 symbols starting from the difference series (we used symbol $'a'$ in case of large decrement; $'b'$: slight decrement; $'c'$: slight increment; $'d'$: huge increment).

We analyzed the frequency of words of length $p = 12$ modulo permutations of the four symbols. That is, any two words of length $p = 12$ are equivalent if they have the same content in symbols $a$, $b$, $c$ and $d$. They were identified by the 4-uple $(n_a, n_b, n_c, n_d)$. This choice is motivated by the specific context where the multi-dimensional series come from: such words of length 12 represent what type of investments occurred over three months, without paying attention to their exact chronological order. This way, it is also easier to get some statistics, since without equivalence there were $4^{12}$ words to look at, while in this case the words are just 2148. Anyway, due to the short length of the series, the number of 12-words used by the 42 brands range from 2 to around 60.

We found that many words were $rare$, that is, they occurred with frequency lower than $1\%$; therefore, we decided to calculate Zipf coefficient only for non-rare words. Of course, a finer analysis should also include which specific words have been used more frequently, but this is not what this example is devoted to.

As a result on the global collection of 42 brands, we selected three categories of diversification (as in section III-C). The brands are said to be highly diversified if $0.8 < \mathcal{D} \leqslant 1$ (they are $64\%$ of the total). If $0 < \mathcal{D} \leqslant 0.8$, then the brands are said to be rich ($14\%$). When $\mathcal{D} \leqslant 0$, they are totally unbalanced ($9\%$).

Since we are dealing with money investments, we also take under consideration the marker relative to the grand total of money invested over the period 1996-2006. Fig. 5 compares the grand total to the Euclidean norm of the entropy vector for the 42 brands in the collection. There is a neat tendency to have higher entropy for huge investments. Notwithstanding, the values of $||H||$ may be wide spread with fixed grand total, especially for intermediate investments.

## V. Final discussion

In this paper we show an analysis of multi-dimensional sparse time series via entropy and statistical linguistic techniques.

Given some phenomenon on which $N$ different measures have been exploited over some time lag, we obtain an $N-$dimensional time series $\mathbb{X} = (X_1, \ldots, X_N)^T$. We have illustrated a way to associate to $\mathbb{X}$ the following markers, which encode the behavior of the series.

- leading component of the complete series $\mathcal{L}(\mathbb{X})$

It refers to the one-dimensional series $X_\mathcal{L}$ in $\mathbb{X}$ whose dynamic is driving the evolution of the overall $N$-dimensional phenomenon.

- trend $\mathcal{T}(\mathbb{X}) = (\mathcal{A}, \alpha)$ w.r.t. $k-$window analysis

It quantifies how much the dynamics has changed in time, in terms of leading component and direction of entropy change.

- diversification $\mathcal{D}(\mathbb{X})$

It formalizes the differentiation in the use of recurrent patterns.

These markers have to be considered as the starting point of further analysis such as classification or clustering of large database of multi-dimensional time series, prediction of future behavior and attribution of new data.

We also present an application to economic data. We deal with measurements of money investments of some business companies in advertising market for different media sources and we point out how to characterize the behavior of each company with respect to the different media, showing a way to label their features.